\documentclass[aps,preprint,epsfig,multicol,nofootinbib,showpacs]{revtex4-1}
\usepackage{graphicx}
\usepackage{amsmath, amsthm, amssymb}
\usepackage{comment}
\usepackage{color}
\begin{document}
\title{Quarantine generated phase transition in epidemic spreading}
\date{\today}
\author{C.~Lagorio,$^{1*}$\footnote[0]{$^*$ The first two authors contributed equally to this paper.} M. Dickison,$^{2*}$ F. Vazquez,$^3$ L. A. Braunstein,$^{1,2}$ P. A. Macri,$^1$ M. V. Migueles,$^1$ S. Havlin,$^4$ and H. E. Stanley$^1$ }
\affiliation{$^1$Instituto de Investigaciones F\'isicas de Mar del Plata (IFIMAR)-Departamento de F\'isica, Facultad de Ciencias Exactas y Naturales, Universidad Nacional de Mar del Plata-CONICET, Funes 3350, (7600) Mar del Plata, Argentina.\\
$^2$Center for Polymer Studies, Physics Department, Boston University, Boston, Massachusetts 02215, USA.\\
$^3$Max-Planck Institute for the Physics of Complex Systems,
Nothnitzer Strasse 38, 01187 Dresden, Germany. \\
$^4$Minerva Center, Department of Physics, Bar Illan University, Ramat Gan, Israel.}
\begin{abstract} We study the critical effect of quarantine on the propagation
of epidemics on an adaptive network of social contacts.  For this purpose, we
analyze the susceptible-infected-recovered (SIR) model in the presence of
quarantine, where susceptible individuals protect themselves by disconnecting
their links to infected neighbors with probability $w$, and reconnecting
them to other susceptible individuals chosen at random.  Starting from a
single infected individual, we show by an analytical approach and
simulations that there is a phase transition at a critical rewiring (quarantine) threshold $w_c$ separating a phase ($w<w_c$) where the disease reaches a large fraction of the population,
from a phase ($w \geq w_c$) where the disease does not spread out.  We find
that in our model the topology of the network strongly affects the size of the
propagation, and that $w_c$ increases with the mean degree and heterogeneity
of the network. We also find that $w_c$ is reduced if we perform a preferential rewiring, in which the rewiring probability is proportional to the degree of infected nodes.
\end{abstract}
\maketitle

\newcommand{\bee}{\begin{eqnarray*}}
\newcommand{\eee}{\end{eqnarray*}}

\newcommand{\be}{\begin{equation}}
\newcommand{\ee}{\end{equation}}
\newcommand{\bc}{\begin{center}}
\newcommand{\ec}{\end{center}}
\newcommand{\bi}{\begin{itemize}}
\newcommand{\ei}{\end{itemize}}
\newcommand{\ba}{\begin{eqnarray}}
\newcommand{\ea}{\end{eqnarray}}
\newcommand{\ie}{{\it i.e.}}
\newcommand{\eg}{{\it e.g.}}
\newcommand{\ignore}[1]{}
\newcommand{\ul}{\underline}

\section{Introduction}

The representation of interactions in a human society as a complex network,
where nodes and links play the respective roles of individuals and their
contacts, has been useful for modeling, studying, and understanding many
problems in epidemiology \cite{Newman, Vespignani-book}. Usually it is assumed
that diseases evolve faster than the topological evolution of the underlying network, so
that the links of the network can be regarded as static. With modern mass media,
however, the presence of an epidemic can be broadcast much faster than
disease propagation. This information will inevitably change the behavior of
the individuals comprising this network as, for example, they try to avoid
contacts with infected people. In this way a feedback loop between the
state of individuals and the topology of the network is formed. Networks that
exhibit such feedback are called adaptive or coevolutionary networks
\cite{Gross, Schwartz, Gross2, Gross3,SanMiguel}.

Public health services are constantly searching for new ways to try to reduce
the spread of diseases. Interventions like vaccination \cite{Anderson_May} or
total closure of workplaces and schools are very effective, but come with a
high economic cost. As a less expensive alternative we examine here the
effectiveness of a
\textquotedblleft quarantine\textquotedblright \hspace{1 pt}strategy,
where healthy people are \textquotedblleft advised\textquotedblright \hspace{1
pt}to avoid contacts with individuals that carry the disease.  That is, a
healthy person has a chance to suppress a contact with an infected neighbor
and form a new tie with another healthy peer (rewiring).  The value of this
rewiring probability could depend, for instance, on the concern that the society has
about the disease and, as suggested above, in a globalized world this concern
will depend on the broadcast news.  The degree of media attention about the
disease is a parameter that could be controlled by public health services.
Indeed, it is known that spontaneous quarantine in the recent H1N1 pandemic
was found to have a large impact in reducing the final size of the epidemic
\cite{H1N1}.

Based on these observations, we propose two strategies for the propagation of
epidemics on an adaptive network, in which individuals alter their local
neighborhoods with  constant quarantine probability $w$ as described above, and
systematically study the effect that the quarantine has on epidemic spreading.
We find a phase transition at a critical threshold $w_c$ above which the
epidemic is stopped from spreading.  We show also how the epidemic spreading
in the presence of quarantine depends on the contagion and recovery parameters.
More importantly, we find that the initial structure of the network plays an
essential role in disease propagation at criticality, unlike in previous related models
\cite{Gross3, Shaw, Zanette, Funk} where results only depend on the average network
connectivity. We also introduce a generalized form of quarantine, where the probability of rewiring is proportional to the degree of the infected nodes, producing a more efficient isolation of the nodes with high degree.
This preferential rewiring is more efficient than the case with $w$ constant.

\section{Analytical approach}
We consider the susceptible-infected-recovered (SIR) epidemic spreading model,
which is well established and accurately describes diseases such as seasonal
influenza, SARS, or AIDS \cite{Anderson_May,Colizza}.  Initially, all nodes
are in the susceptible state ($s$), with one node chosen at random (seed) in the
infected state ($i$). In the first strategy, strategy $A$, at each time unit, every infected node $i$ in the network
tries to transmit the disease to each susceptible neighbor $s$ with infection
probability $\beta$.  If $s$ does not get infected, then with rewiring
probability $w$ it disconnects its link from $i$ and reconnects it to another
randomly chosen susceptible node, different from its present neighbors.  Thus,
the rewiring probability $w$ measures how fast susceptible nodes react to the
disease (the quarantine probability).  Infected nodes recover ($r$) after a fixed recovery
time $t_R$ since they first became infected, remaining in the recovered state
forever.

To estimate $w_c$ we start by assuming that the network has a tree structure
\cite{tree}, so that the disease spreads out from the seed and reaches a
susceptible node $s$ through only one of its neighbors $i$.  Then, if $i$
becomes infected at time $t_0$, the probability that $s$ becomes infected by
$i$ at time $t_0+n$, with $n=1,2,..t_R$, is $\beta (1-\beta)^{n-1}
(1-w)^{n-1}$.  This is the probability that $s$ has neither become infected
nor disconnected from $i$ up to time $t_0+n-1$, times the probability that $i$
succeeds in transmitting the disease to $s$ at time $t_0+n$.  Therefore, the
overall probability that $s$ becomes infected before $i$ recovers is given by
the sum
\begin{eqnarray}
T_{\beta,w}^{A} &\equiv& \sum_{n=1}^{t_R} \beta (1-\beta)^{n-1} (1-w)^{n-1}
\nonumber \\ &=& \frac{\beta \Bigl\{1-\left[(1-\beta) (1-w) \right]^{t_R}
\Bigr\}} {1-(1-\beta)(1-w)}.
\label{T}
\end{eqnarray}

This expression for the transmissibility $T_{\beta,w}^A$ is equivalent to the
corresponding expression in the standard SIR model $\sum_{n=1}^{n=t_R} \beta
(1-\beta)^{n-1}$ \cite{Newman}, but with a non-infection probability
$(1-\beta)(1-w)$, instead of $(1-\beta)$.  When $w=0$, Eq.~(\ref{T})
reproduces the known value for the transmissibility $T =1-(1-\beta)^{t_R}$ for
the SIR model on static networks \cite{Newman}.  In this formulation $w$ plays
the role of a control parameter of the transmissibility: for fixed values of
$t_R$ and $\beta$, $T_{\beta,w}^A$ can be reduced by increasing $w$.  By
reducing $T_{\beta,w}^A$ we can go from a regime in which the epidemic spreads
over the population (epidemic phase) to another regime where the disease
cannot spread (disease-free phase).  The transition from the disease-free
phase to the epidemic phase corresponds to the average number of secondary
infections per infected node becoming larger than one, allowing the long-term
survival of the disease and thus ensuring the epidemic spreads to a large
fraction of the population. In our problem, the expected number of susceptible
neighbors that a node has when it just becomes infected is given by $\kappa -1$,
where  $\kappa -1 \equiv \langle k^2 \rangle / \langle k \rangle -1$ is called the \emph{branching factor}, and $\langle k \rangle$ and $\langle k^2 \rangle$
are the first and second moments, respectively, of the degree distribution
$P(k)$. Since $T_{\beta,w}^A$ is the overall probability to infect a neighbor, the mean number of secondary
infections per infected node is

\begin{equation}
N_I^A(w) = (\kappa -1) T_{\beta,w}^A.
\end{equation}

The infection will die out if each infected node does not spawn on average at least one
replacement, so for a very large system, the critical point is given by the
relation $(\kappa -1) T_{\beta,w_c}^{A}=1$, or
\begin{equation}
\frac{ (\kappa -1) \beta \Bigl\{1-\left[(1-\beta) (1-w_c) \right]^{t_R}
\Bigr\}} {1-(1-\beta)(1-w_c)} = 1.
\label{transition}
\end{equation}

The transition between free-disease phase and epidemic phase is analogous to the \emph{static link percolation} problem \cite{Havlin_Book}, in which each link in a network is occupied with probability $p$ and empty with probability
$q=(1-p)$. When $p$ becomes smaller than a percolation threshold $p_c$, the
giant connected component disappears. In general, $p_c$ depends on the size of the network $N$\cite{Wu}, but in the
thermodynamic limit it can be expressed as $p_c = 1/(\kappa -1)$ \cite{Cohen}. Identifying $p$ with the
transmissibility $T_{\beta,w}^A$, and using the relation between $p_c$ and $\kappa$  we find that on the epidemic/disease-free
transition line:
\begin{equation}
T_{\beta,w_c}^A = p_c.
\label{T-pc}
\end{equation}

This result shows that the transition point depends on the initial topology of
the network through the moments of the degree distribution, as we shall
confirm via simulation.

A better strategy would be try to avoid the contact between susceptible and infected individuals before the attempt to infection, i.e., only avoid the contact when you know an individual is sick. In this second strategy, strategy B, at each time unit every susceptible node attached to an infected node disconnects its link from $i$ with probability $w$ and reconnects it to another randomly chosen susceptible node, different from its present neighbors. If $s$ does not rewire its link, the infected node $i$ tries to infect it with infection probability $\beta$.
Notice that $N_I^B(w)=(1-w) N_I^A(w)$.
For this strategy Eq.(\ref{T}) is replaced by

\begin{eqnarray}
T_{\beta,w}^{B}&\equiv &\frac{\beta (1-w) \Bigl\{1-\left[(1-\beta) (1-w) \right]^{t_R} \;,
\Bigr\}} {1-(1-\beta)(1-w)}.
\label{T1}
\end{eqnarray}
Comparing Eq.(\ref{T}) and Eq.(\ref{T1}) we see that  $T_{\beta,w}^{B}<T_{\beta,w}^{A}$ for identical values of $\beta$ and $w$.  The new condition for disease to die in strategy B out follows from Eq.(\ref{T1})

\begin{equation}
\frac{ (\kappa -1) \beta (1-w_c) \Bigl\{1-\left[(1-\beta) (1-w_c) \right]^{t_R}
\Bigr\}} {1-(1-\beta)(1-w_c)} = 1.
\label{transition1}
\end{equation}

The two strategies represent different scenarios depending on the knowledge of the state of infection of the nodes. For strategy A, susceptible nodes have no information about the state of their neighbors until they are in physical contact. In contrast, for strategy B susceptible nodes know the state of their neighbors before they are in physical contact. We will show that this difference in the knowledge of the states of the nodes results in strategy B being more effective at stopping epidemic spreading.

\begin{figure}[h]
\includegraphics[width=8cm,height=6cm]{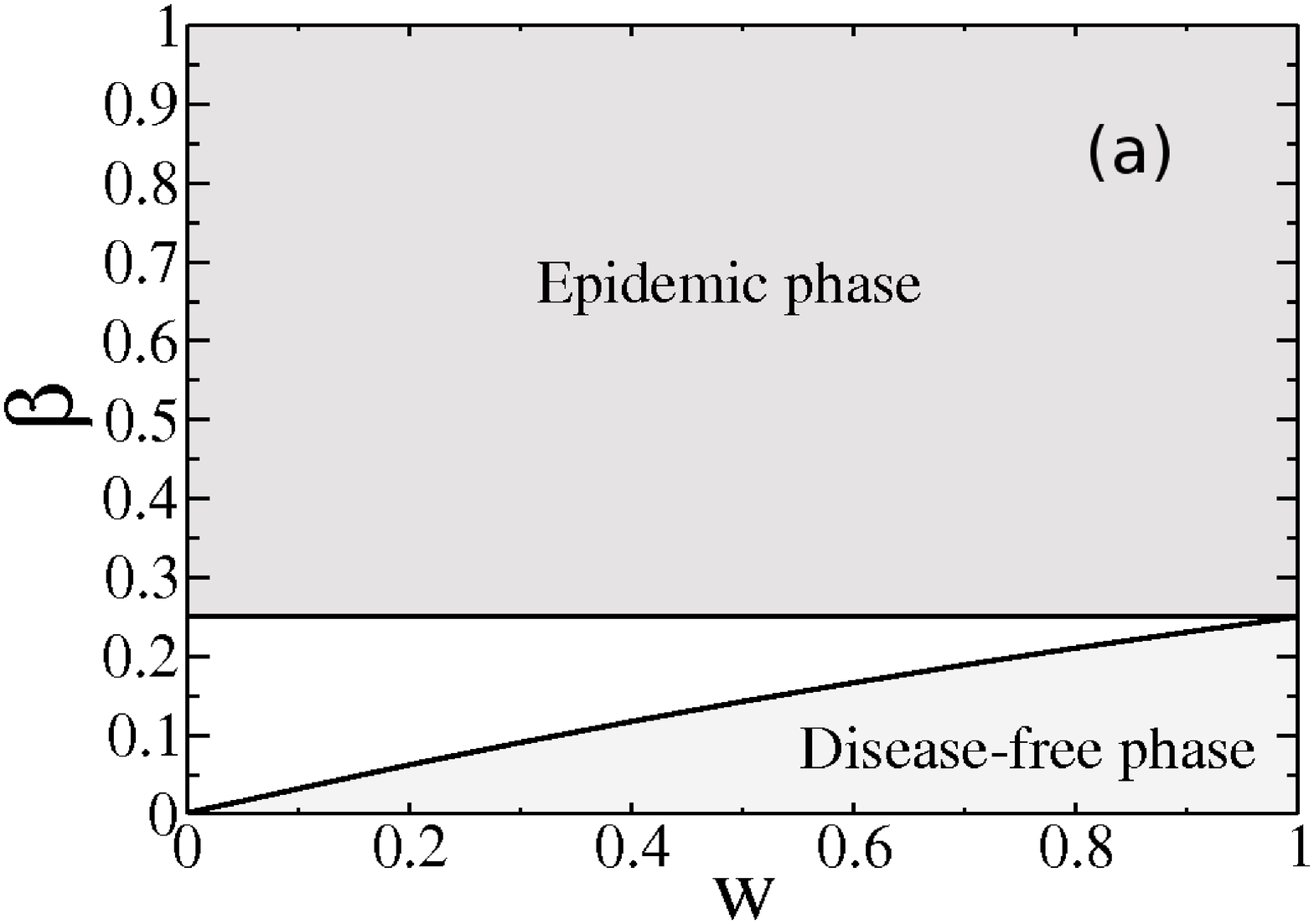}
\includegraphics[width=8cm,height=6cm]{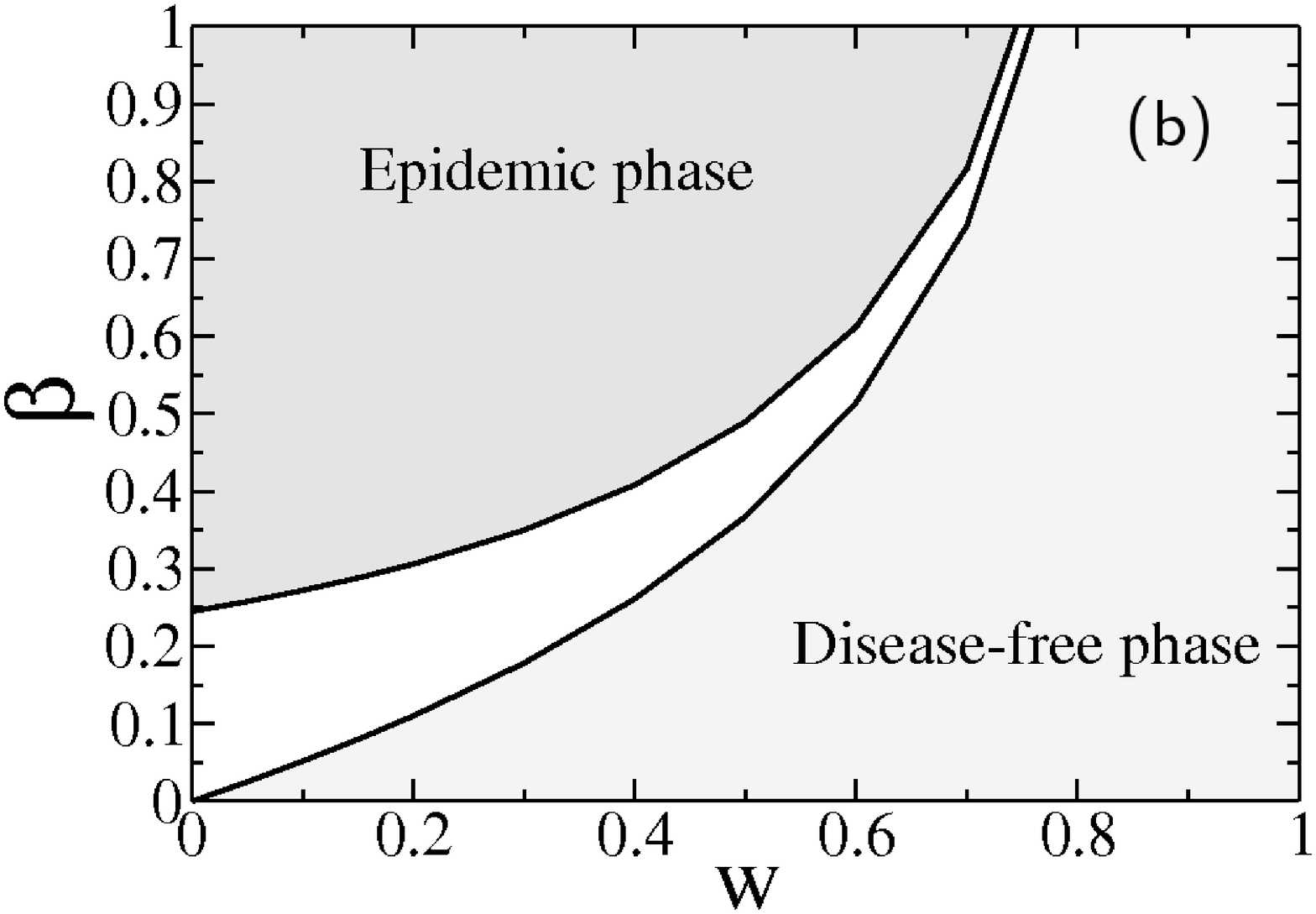}
\caption{$\beta$-$w$ phase diagram for $p_c=0.25$. The
curves correspond to $t_R=1$ (upper) and $t_R=\infty$ (lower). The dark gray
region will always be an epidemic phase, and the light gray always a
disease-free phase: (a) strategy A, (b) strategy B.
\label{f.00}}
\end{figure}

Fig.~\ref{f.00} shows phase diagrams  in the $w-\beta$ plane for strategy A and strategy B, obtained by the
numerical solutions of Eq.~(\ref{transition}) and Eq.~(\ref{transition1}) respectively, for a network with $\kappa = 5$
($p_c = 0.25$).  Two phases emerge: the epidemic phase for $T_{\beta,w}^{A,B}>p_c$,
and the disease-free phase for $T_{\beta,w}^{A,B} \leq p_c$.  The location of the
critical line separating the two phases depends on the recovery time
$t_R$. Fig.~\ref{f.00}(a) illustrates the two limiting cases for strategy A:
$t_R=1$ (upper curve) and $t_R=\infty$ (lower curve), which are given by the
expressions $\beta(t_R=1)=1/(\kappa-1)$ and
$\beta(t_R=\infty)=w/(\kappa+w-2)$.  Therefore, the pairs of $(w,\beta)$
values in the region above the curve $\beta(t_R=1)$ are always in the epidemic
phase, and below the curve $\beta(t_R=\infty)$ are always in the disease-free
phase. A striking consequence is that if $\beta$ is larger than the
percolation threshold $1/(\kappa -1)$, the propagation of an epidemic cannot
be stopped, even with the largest rewiring probability $w=1$.  Thus,
strategy A is not an efficient mechanism to control an epidemic when $\beta$
is higher than $p_c$.

For strategy B, Fig.~\ref{f.00}(b) illustrates the two limiting cases
$t_R=1$ (upper curve) and $t_R=\infty$ (lower curve), which are given by the
expressions $\beta(t_R=1)=1/(\kappa-1)(1-w)$ and
$\beta(t_R=\infty)=w/(\kappa -\kappa w + 2w - 2)$. In contrast to strategy A above, if $\kappa$ does not diverge, i.e., the network has finite $p_c$, the epidemic can always be stopped, even for $\beta \rightarrow 1$. Notice that the maximum value of $w$ needed to stop a epidemic is $w_c=(\kappa - 2)/(\kappa - 1)=1-p_c$.

Our theoretical predictions illustrate a novel feature about the
dynamics of adaptive networks in SIR models.  While previous adaptive SIS and SIRS models have
transition values that depend only on the average connectivity of the network
$\langle k \rangle$, and are independent of the heterogeneity or
structural correlations of the initial topology\cite{Gross}, our SIR model predicts dependence on the topological structure through the higher order moments of the degree distribution.

\section{Simulation Results}

Our analytical approach predicts, through Eq.~(\ref{transition}) and Eq.~(\ref{transition1}), a dependence
of $w_c$ on the initial network.  In order to test and explore this
dependence, we performed extensive numerical simulations of our model starting
from different topologies.  We first used Erd\"{o}s-R\'enyi (ER) networks with Poissonian degree
distribution $P(k)=e^{-\langle k \rangle} \langle k \rangle^k / k!$.
Even though these types of homogeneous networks are common in nature, many
real social networks are well represented by heterogeneous networks. Thus, we also used
finite scale-free (SF) networks with degree distribution
$ P(k)=k^{-\lambda}\exp(-k/K)/\mbox{Li}_{\lambda}(e^{-1/K})$,
where $K$ is the degree cutoff.  This distribution represents networks with a finite threshold $p_c$, and appears in a variety of real-world
networks \cite{Aiello, Bollobas}. We only consider epidemic propagation on the largest connected
cluster of the network, the giant component (GC).

\begin{figure}[h]
\includegraphics[width=8cm,height=6cm]{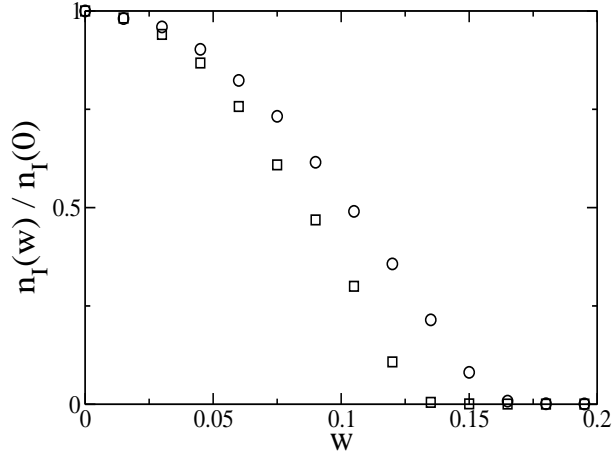}
\caption{Plot of $n_I(w)/n_I(0)$ as a function of $w$ for strategy A ($\bigcirc$), and strategy B ($\Box$), for a ER network with $\langle k \rangle _{GC} =4.07$, $N=10^4$ and $10^4$ realizations. As we predicted previously, strategy B is better than strategy A, shown by $w_c^B <w_c^A$.
\label{f.comp}}
\end{figure}

In Fig.~\ref{f.comp} we compare both strategies for ER networks. We plot as a function of $w$, the average fraction of
infected nodes $n_I(w)= N_I(w)/N_{GC}$, where $N_{GC}$ is the size of the giant
component, on a ER network.  This fraction is normalized
by the corresponding fraction on an identical fixed network, i.e., for the SIR
model without quarantine, $n_I(0)$. We can see that $w_c$ for strategy B is lower than $w_c$ for strategy A, as expected. This relation will hold for any topology with the same $\kappa$ (see Eq.(\ref{transition}) and Eq.(\ref{transition1})). Since strategy B is more effective, from here to the end of the paper we will show only simulation results for strategy B.

\hspace{2cm}

\begin{figure}[h]
\includegraphics[width=7.5cm,height=5cm]{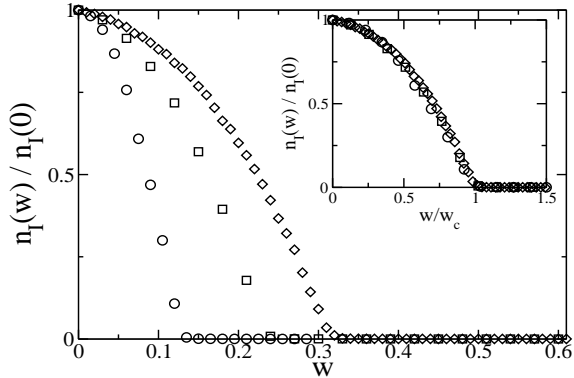}
\caption{Fraction of nodes infected as a function of the applied quarantine parameter $w$ divided by the fraction of infected nodes at $w=0$ for three different infection probabilities
$ \beta$ in an ER network. ($\bigcirc$)
$\beta=0.05$, ($\Box$) $\beta=0.1$ and ($\Diamond$) $\beta=0.15$. Above a threshold value of $w$, no finite fraction
of the network becomes infected.  (Inset) Data with $w$ rescaled by the appropriate $w_c$ calculated from Eq.~(\ref{transition1}).
The curves collapse very well, showing universal behavior and good
agreement with the theory. In all simulations, $\langle k \rangle_{GC} =4.07$ ($\langle k \rangle$ = 4 over all nodes),
$N=10^4$ and averages are over $10^4$ realizations.\label{f.1}}
\end{figure}

Fig.~\ref{f.1} shows, $n_I(w)/n_I(0)$ vs $w$ for different $\beta$. In this figure, we can observe the strong
effect of quarantine and the critical threshold $w_c$ above which $n_I(w)$
approaches zero.  Thus, in the disease-free phase ($w \geq w_c)$ only a small
number of individuals get infected and the disease quickly dies out.  We
observe, as expected, that the values of $w_c$ increase with $\beta$.  This behavior matches
the phase diagram of Fig.~\ref{f.00} (b), where the critical line has a positive
slope. Scaling the horizontal axis by the values of $w_c$ obtained by numerically solving
Eq.~(\ref{transition1}) collapses the curves, showing a excellent agreement between
theory and simulations, as well as a scaling behavior of the form $n_I (w) =
n_I(0) f\left(w/w_c \right)$, as shown in the inset of Fig.~\ref{f.1}.

\hspace{2cm}

\begin{figure}[h]
\includegraphics[width=7.5cm,height=5cm]{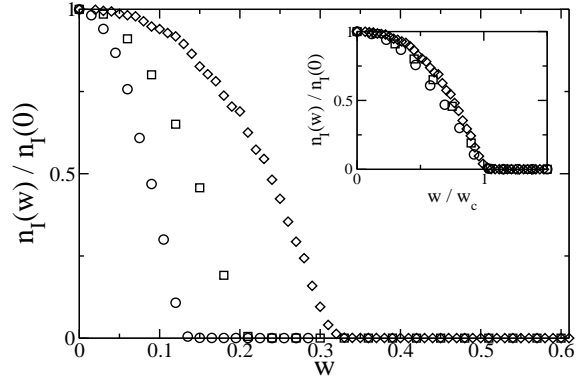}
\caption{Fraction of nodes infected as a function of the applied quarantine parameter $w$ divided by the fraction of infected nodes at $w=0$ for ER networks with three different $\langle k
\rangle$: $\langle k \rangle_{GC}=4.07$ ($\bigcirc$), $\langle k \rangle_{GC}=6.015$ ($\Box$) , $\langle k \rangle_{GC}=10$ ($\Diamond$). Again, while below $w_c$ a
finite fraction of the network become infected above $w_c$ the epidemic vanishes. (Inset) Data rescaled by $w_c$ calculated
from Eq(\ref{transition1}) . In all simulations, $\beta=0.05 $, $N=10^4$ and averages are over $10^4$
realizations.\label{f.2}}
\end{figure}

We also explored the dependence of $w_c$ on the connectivity of the network,
by computing $n_I(w)/n_I(0)$ vs $w$ for different $\langle k \rangle$ (see
Fig.~\ref{f.2}).  We observe that $w_c$ increases with $\langle k \rangle$,
due to the fact that propagation is facilitated by having more neighbors,
as in the original SIR dynamics.  The inset of Fig.~\ref{f.2} shows the collapse of all curves.  Again, the good agreement
between theory and simulation confirms that Eq.~(\ref{transition1}) is a valid
expression of the transition point for the adaptive SIR model on ER networks.

Given that the second moment, and therefore $\kappa$, is large in heterogeneous networks, the critical value $w_c$ turns out to be larger in heterogeneous networks than in homogeneous networks with the same mean degree $\langle k \rangle$ and size $N$, as seen by considering
Eq.~(\ref{transition1}) in the $t_R \gg 1$ limit, where
\begin{equation}
w_c \simeq \frac{1}
{\frac{1}{\beta (\kappa -2)}+1} .
\label{w_K}
\end{equation}

Since $\kappa$ for heterogeneous networks is much bigger than in homogeneous networks we expect that $w_c$ increases as the heterogeneity increases with larger $K$. For $\kappa \rightarrow \infty$ we expect that $w_c \to 1$ and that the transition will eventually
disappear for large heterogeneous networks.

\begin{figure}[h]
\includegraphics[width=7.5cm,height=5cm]{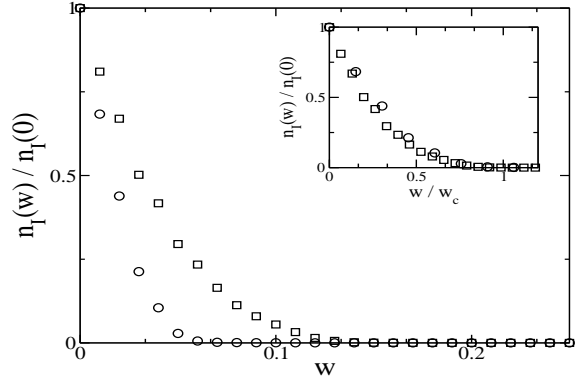}
\caption{ Fraction of infected nodes still infected as a function of $w$ for $\beta=0.05$, $t_R=20$
on an SF network with $K=10$ ($\bigcirc$) and $K=20$ ($\Box$), $\lambda = 2.1$ with $N=10^5$.  $p_c\cong0.375$ for $K=10$ and  $p_c\cong0.215$ for $K=20$. (Inset) Rescaled by $w_c$.  The result shows an
excellent agreement between the theory and the simulation.
\label{f.4}}
\end{figure}

In Fig.~\ref{f.4}, we show simulations on scale-free networks for different values
of $K$\cite{Pc_N}. In good agreement with our predictions from Eq.~(\ref{w_K}), as $\kappa$ increases with $K$, $w_c$ increases. In the inset we rescale by $w_c$ obtained from Eq.~(\ref{transition1}), and find good collapse
(Fig.~\ref{f.4}(inset)), confirming the general validity of Eq.~(\ref{transition1}) for heterogeneous SF networks.

In very heterogeneous networks it is well known that due to high degree nodes, propagation processes are very difficult to stop \cite{Havlin_Book, Vespignani}. Thus, a constant probability of rewiring is not effective for controlling epidemics in these networks, because such a strategy assumes all nodes have the same importance in an epidemic, ignoring the function of higher degree nodes as superspreaders. It is also well known that removing the high degree nodes of the network (targeted percolation) \cite{Havlin_Book, Barabasi-Nature, Cohen2001} is a more efficient method to stop propagation processes than random removal. This type of strategy is expected to be superior, but requires global information about the network.

To test this prediction, we propose a new strategy of type B where $w$ depends on the degree $k$ of the infected node, with $w_k$ given by the general form
\begin {equation}
w_k\equiv w_k(\alpha)=\gamma k^\alpha ~,
\end {equation}
where $\gamma$ is a constant that controls the highest possible value of $w_k$, and it is equal or smaller than $k^{-\alpha}_{max}$, where $k_{max}$ is the largest degree of the network. For $\alpha = 0$ and $\gamma$ $\in$ [0,1] we recover the results for a constant value of $w$, with $w = \gamma$. For $\alpha > 0$ and $\gamma  = k^{-\alpha}_{max}$ the rewiring increases with the degree $k$ of the infected node, and decreases with increasing $\alpha$. In the limit of $\alpha \rightarrow \infty$ the rewiring process is equivalent to a targeted rewiring where only the links of the highest connected node(s) are rewired.  To compare the cases with $\alpha = 0$ and $\alpha>0$, we use the average $\langle w \rangle$ over the network, choosing $\alpha$ for the targeted case such that  $\langle w \rangle$ is equal to $w$ in the uniform case.

\begin{figure}[h]
\includegraphics[width=7.5cm,height=5cm]{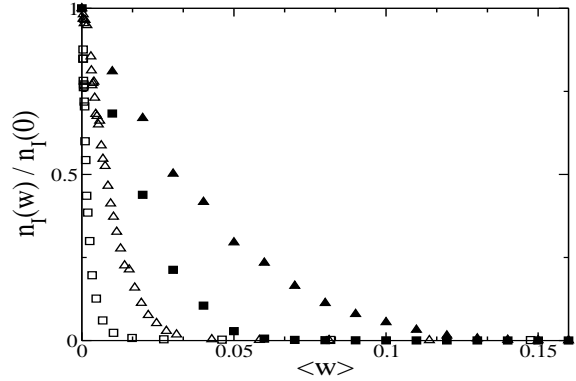}
\caption{ Fraction of infected nodes still infected as a function of $\langle w \rangle$ for $\beta=0.05$, $t_R=20$
on an SF network, $\lambda = 2.1$ , $N=10^4$ and $K=10 , 20$. For general strategy B with $\alpha=0$, ($K=10$ ($\blacksquare$) and $K=20$ ($\blacktriangle$)), and with $\alpha>0$, ($K=10$ ($\Box$) and $K=20$ ($\vartriangle$), each $\langle w \rangle$ represents a different $\alpha$. \label{f.3}}
\end{figure}

In Fig.~\ref{f.3} we plot $n_I(w)/n_I(0)$ as a function of $\langle w \rangle$ for SF networks with $K=10$ and $K=20$. As expected, $w_c$ for $\alpha>0$ is lower than for $\alpha=0$. The preferential rewiring reduces the value of $w_c$ because quarantine is more effective at isolating the superspreaders.

\section{Summary and conclusions}

We have introduced and studied two strategies for the propagation of epidemics on evolving networks of social contacts.  The states of the individuals are
changed according to SIR dynamics, while the network evolves according to
a quarantine mechanism based on local information, in which susceptible
individuals replace their infected neighbors by other susceptible peers with probability $w$.  We
demonstrated by an analytical approach, and confirmed by numerical simulations,
that the size of the epidemics can be largely reduced by increasing the probability
of rewiring, and that the propagation can be eventually stopped by using
high enough values of $w$.  In other words, quarantine is an effective way to
halt the appearance of epidemics that would otherwise emerge in the case of a
static network. For strategy A, when the infection probabliy is larger than
a threshold the quarantine mechanism is not effective any more and
disease propagation becomes unavoidable.  This is because the quarantine model only breaks contact after allowing for a chance of infection,  thus for high enough infection probability, the spreading is irreversible, even with a rewiring probability of unity.
For strategy B, the quarantine mechanism is effective in homogeneous networks and finite heterogeneous networks, even for infection probability equal to unity. The transition disappears only for large, very heterogeneous networks.
In these scale-free networks, the critical rewiring probability becomes very large and the transition eventually
disappears for large enough systems, thus only the epidemic phase is observed.
This is likely due to the presence of individuals with very large
connectivity that can spread the disease over a large fraction of the
population, even for small infection probabilities.  To confirm this, we introduced a generalized form of strategy B, where the quarantine depends on the degree of the infected nodes. This preferential rewiring isolates the superspreaders more efficiently, reducing $w_c$ and preserving a finite transition even for a scale-free networks. Lastly, in SIR dynamics the final frozen state, where everybody is either recovered or susceptible, is reached rather quickly ---in a time of order $\ln N$ according to our simulations--- and thus an SIR network does not evolve much, so the initial topology is preserved during the entire evolution of the system.  In contrast, in adaptive models with SIS or SIRS dynamics the system evolves for a very long time in the epidemic phase ---times that grow exponentially larger with $N$--- thus the network moves towards a stationary topology similar to an  Erd\"{o}s-R\'enyi network, independent of the initial topology, and rendering the initial topology irrelevant. Therefore, unlike other adaptive network models of epidemic spreading, in our model the epidemic threshold has a strong correlation with the topology of the network, which remains relatively unchanged at criticality, and the social structure of connections when epidemics begin to propagate is crucial in the state of the final outbreak.

\begin{acknowledgments}
 CL, LAB, PAM, MVM thank UNMdP and FONCYT-PICT 2008/0293 for financial support. MD, SH and HES thank the DTRA, European EPIWORK project, and the ONR for financial support, SH thanks also the Israel Science Foundation.
\end{acknowledgments}

\end{document}